\documentclass[twocolumn,showpacs,fleqn,nobibnotes]{revtex4}
\usepackage{amsmath}
\usepackage{graphicx}
\usepackage{float}
\usepackage{subfigure}
\def\lsim{\raise0.3ex\hbox{$<$\kern-0.75em\raise-1.1ex\hbox{$\sim$}}}
\def\gsim{\raise0.3ex\hbox{$>$\kern-0.75em\raise-1.1ex\hbox{$\sim$}}}

\begin{document}
\newcommand\ie {{\it i.e.}}
\newcommand\eg {{\it e.g.}}
\newcommand\etc{{\it etc.}}
\newcommand\cf {{\it cf.}}
\newcommand\etal {{\it et al.}}
\newcommand{\be}{\begin{eqnarray}}
\newcommand{\ee}{\end{eqnarray}}
\newcommand{\jp}{$ J/ \psi $}
\newcommand{\pp}{$ \psi^{ \prime} $}
\newcommand{\ppp}{$ \psi^{ \prime \prime } $}
\newcommand{\dd}[2]{$ #1 \overline #2 $}
\newcommand\noi {\noindent}

\title{Investigating gluino production at the LHC}
\pacs{12.60.Jv; 14.80.Ly, 13.85.Qk}
\author{C. Brenner Mariotto $^{a}$ and M.C. Rodriguez $^{a}$}

\affiliation{
$^a$ Departamento de F\'{\i}sica, Funda\c{c}\~ao Universidade Federal do Rio Grande \\
Caixa Postal 474, CEP 96201-900, Rio Grande, RS, Brazil
}

\begin{abstract}
Gluinos are expected to be one of the most massive sparticles 
(supersymmetric partners of usual particles) which constitute the Minimal 
Supersymmetric Standard Model (MSSM). The gluinos are the partners of the 
gluons and they are color octet fermions, due this fact they can not mix with 
the other particles. Therefore in several scenarios, given at SPS convention, 
they are the most massive particles and their nature is a Majorana fermion. 
Therefore their production is only feasible at a very energetic machine 
such as the Large Hadron Collider (LHC). Being the fermion partners of the 
gluons, their role and interactions are directly related with the properties 
of the supersymmetric QCD (sQCD). 
We review the mechanisms for producing gluinos at 
the LHC and investigate the total cross section and differential distributions, 
making an analysis of their uncertainties, such as the gluino and squark
masses, as obtained in several scenarios, commenting on the possibilities of
discriminating among them.

\end{abstract}

\maketitle

Although the Standard Model (SM) \cite{sg}, based on the gauge symmetry 
$SU(3)_{c}\otimes SU(2)_{L}\otimes U(1)_{Y}$ describes the observed properties
of charged leptons and quarks it is not the ultimate theory. 
However, the necessity to go beyond it, from the 
experimental point of view, comes at the moment only from neutrino 
data.  If neutrinos are massive then new physics beyond the SM is needed.

Although the SM provides a correct description of virtually all known microphysical 
nongravitacional phenomena, there are a number of theoretical and phenomenological issues that the SM fails to address 
adequately \cite{Chung:2003fi}:
\begin{itemize}
\item Hierarchy problem;
\item Electroweak symmetry breaking (EWSB);
\item Gauge coupling unification.
\end{itemize}
The main sucess of supersymmetry (SUSY) is in solving the problems listed above.

SUSY has also made several correct predictions \cite{Chung:2003fi}:
\begin{itemize}
\item SUSY predicted in the early 1980s that the top quark would be heavy;
\item SUSY GUT theories with a high fundamental scale accurately predicted the present experimental value of $\sin^{2} \theta_{W}$ 
before it was mesured;
\item SUSY requires a light Higgs boson to exist.
\end{itemize}
Together these success provide powerful indirect evidence that low energy SUSY is indeed part of correct description of nature.

Certainly the most popular extension of the SM is its supersymmetric counterpart 
called Minimal Supersymmetric Standard Model (MSSM) \cite{mssm}. The main motivation to 
study this models, is that it provides a solution to the hierarchy 
problem by protecting the electroweak scale from large radiative corrections \cite{INO82a,INO82b}. Hence the mass square
 of the lightest real scalar boson has an upper bound given by 
\begin{equation}
M^{2}_{h}\leq (M^{2}_{Z}+\epsilon^{2})\;{\rm GeV}^{2}
\label{ub1}
\end{equation}
where $M^{2}_{Z}$ is the Z mass. Therefore the CP even, light Higgs $h$, is expected lighter than Z at tree level ($\epsilon=0$).
However, radiative corrections rise it to 130 GeV~\cite{haber2}.

In the MSSM \cite{mssm}, the gauge group is $SU(3)_{C}\otimes
SU(2)_{L}\otimes U(1)_{Y}$. The particle content of this model consists in associate to every
known quark and lepton a new scalar superpartner to form a chiral supermultiplet. Similarly, we
group a gauge fermion (gaugino) with each of the gauge bosons of the standard model to form a
vector multiplet. In the scalar sector, we need to introduce two Higgs scalars and also their
supersymmetric partners known as Higgsinos. We also need to impose a new global $U(1)$ invariance 
usually called $R$-invariance, to get interactions that conserve both lepton and
baryon number (invariance).

Other
 very popular extensions of SM are Left-Right symmetric theories \cite{ps74}, 
which attribute the 
observed parity asymmetry in the weak interactions to the spontaneous 
breakdown of Left-Right symmetry, i.e. generalized parity transformations. It 
is characterized by a number of interesting and
important features \cite{melfo1}: 

\begin{enumerate}
\item it incorporates Left-Right (LR) symmetry  which leads naturally to the
spontaneous breaking of parity and charge conjugation;
 
\item incorporates a see-saw mechanism for small neutrino masses.

\end{enumerate}

On the technical side, the left-right symmetric model has a problem similar to
that in the SM: the masses of the fundamental Higgs scalars diverge
quadratically. As in the SM, the Supersymmetric Left-Right Model (SUSYLR) can be used to stabilize the scalar
masses and cure this hierarchy problem.

Another,
 maybe more important {\it raison d'etre} for SUSYLR 
models is the fact that they lead naturally to R-parity conservation  \cite{melfo2}.
Namely, Left-Right models contain a $B-L$ gauge symmetry, which allows
for this possibility \cite{m86}. All that is  needed is that one uses
a version  of the theory that incorporates a see-saw mechanism \cite{seesaw1}
at the renormalizable level.

The supersymmetric extension of left-right models \cite{susylr,doublet} is
based on the gauge group $SU(3)_{C}\otimes SU(2)_{L}\otimes SU(2)_{R}\otimes
U(1)_{B-L}$. On the literature there are two different SUSYLR models. They differ in their
$SU(2)_{R}$ breaking fields: one uses $SU(2)_{R}$ triplets \cite{susylr} (SUSYLRT) and the
other $SU(2)_{R}$ doublets \cite{doublet} (SUSYLRD). Since we are interested in studying only the 
strong sector, which is the same in both models, the results we are presenting here hold in both models.

As a result of a more detailed study, we have shown that the Feynman 
rules of the 
strong sector are the same in both MSSM and SUSYLR models \cite{BR}. The relevant 
Feynman rules for the gluino production are:

- Gluino-Gluino-Gluon: $-g_{s}f^{bac}\,\,;$

- Quark-Quark-Gluon: $- \imath g_{s} T^{a}_{rs} \gamma^{m}\,\,\,$ (usual QCD);

- Squark-Squark-Gluon: $- \imath g_{s}T^{a}_{rs}(k_i^{} + k_j^{})^{m}
\,\,$, where $k_{i,j}$ are the momentum of the incoming and outcoming squarks, respectively;

- Quark-Squark-Gluino: $- \imath \sqrt{2}g_{s}(LT^{a}_{rs}-RT^{a}_{rs})
\,\,\,\,\,\,\,$, where $L=\frac{1}{2}(1-\gamma_5)\,\,,\,\,\,\, R=\frac{1}{2}(1+\gamma_5)\,.\,\,\,\,$

The ``Snowmass Points and Slopes'' (SPS) \cite{sps1} are a set of benchmark points
and parameter lines in the MSSM parameter space corresponding to different scenarios in the search
for Supersymmetry at present and future experiments. The aim of this convention
is reconstructing the fundamental supersymmetric theory, and its breaking
mechanism, from the data. 
The points SPS 1-6 are Minimal Supergravity (mSUGRA) model, 
SPS 7-8 are gauge-mediated symmetry breaking (GMSB) model, and SPS 9 are 
anomaly-mediated symmetry breaking (mAMSB) model (\cite{sps1,sps2,sps}). 
Each set of parameters leads to different 
gluino and squark masses, wich are the only relevant parameters in our study, 
and are shown in Tab.(I).

\begin{table}[hb]
\renewcommand{\arraystretch}{1.10}
\begin{center}
\normalsize
\begin{tabular}{|c|c|c|}
\hline
\hline
Scenario & $m_{\tilde{g}}\, (GeV)$ & $m_{\tilde{q}}\, (GeV)$ \\
\hline
\hline
 SPS1a & 595.2  & 539.9 \\
 SPS1b & 916.1  & 836.2 \\
 SPS2 & 784.4  & 1533.6 \\
 SPS3 & 914.3  & 818.3 \\
 SPS4 & 721.0  & 732.2 \\
 SPS5 & 710.3  & 643.9 \\
 SPS6 & 708.5  & 641.3 \\
 SPS7 & 926.0  & 861.3 \\
 SPS8 & 820.5  & 1081.6 \\
 SPS9 & 1275.2  & 1219.2 \\
  \hline
\hline
\end{tabular}
\caption{The values of the masses of gluinos and squarks in the SPS scenarios.}
\end{center}
\label{tab1} 
\end{table}

Gluino and squark production at hadron colliders occurs dominantly via strong interactions. Thus, their production rate 
may be expected to be considerably larger than for sparticles with just electroweak interactions whose 
production was widely studied in the literature \cite{tata,dress}. Since the Feynman rules of the 
strong sector are the same in both MSSM and SUSYLR models, the diagrams that contribute to the gluino production 
are the same in both models. 

In the present contribution we study the gluino production in $pp$
collisions at LHC energies. 
To make a consistent comparison and for sake of simplicity, we restrict ourselves to leading-order
(LO) accuracy, 
where the partonic cross-sections for the production of squarks and gluinos in hadron
collisions were calculated at the Born level already quite some time ago \cite{Dawson}. 
The corresponding NLO calculation has already been done for the MSSM case \cite{Zerwas}, and the
impact of the higher order terms is mainly on the normalization of the cross
section, which could be
taken in to account here by introducing a K factor in the results here obtained \cite{Zerwas}. 

\begin{figure}[t]
\begin{center}
\includegraphics[scale=0.14]{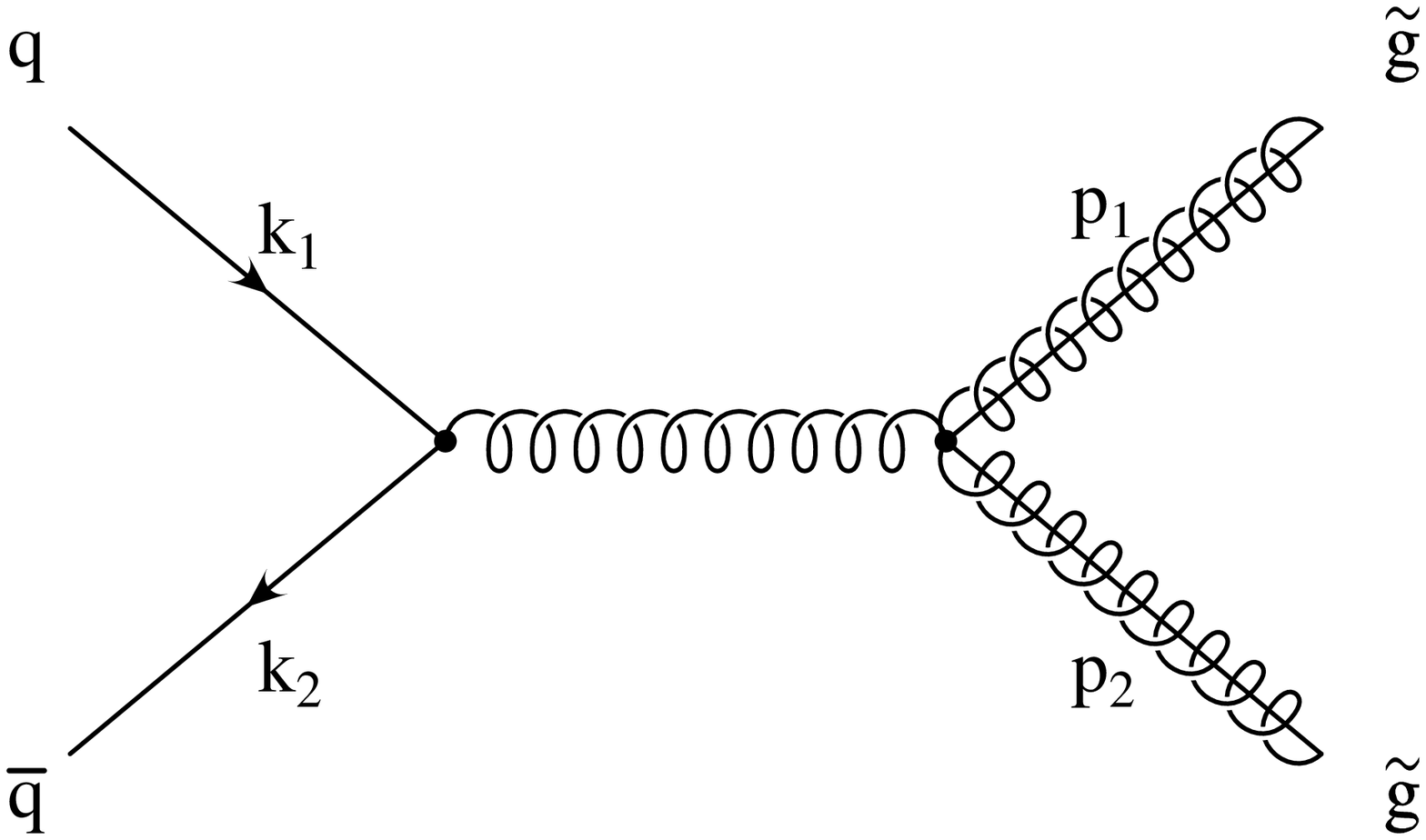}
\includegraphics[scale=0.14]{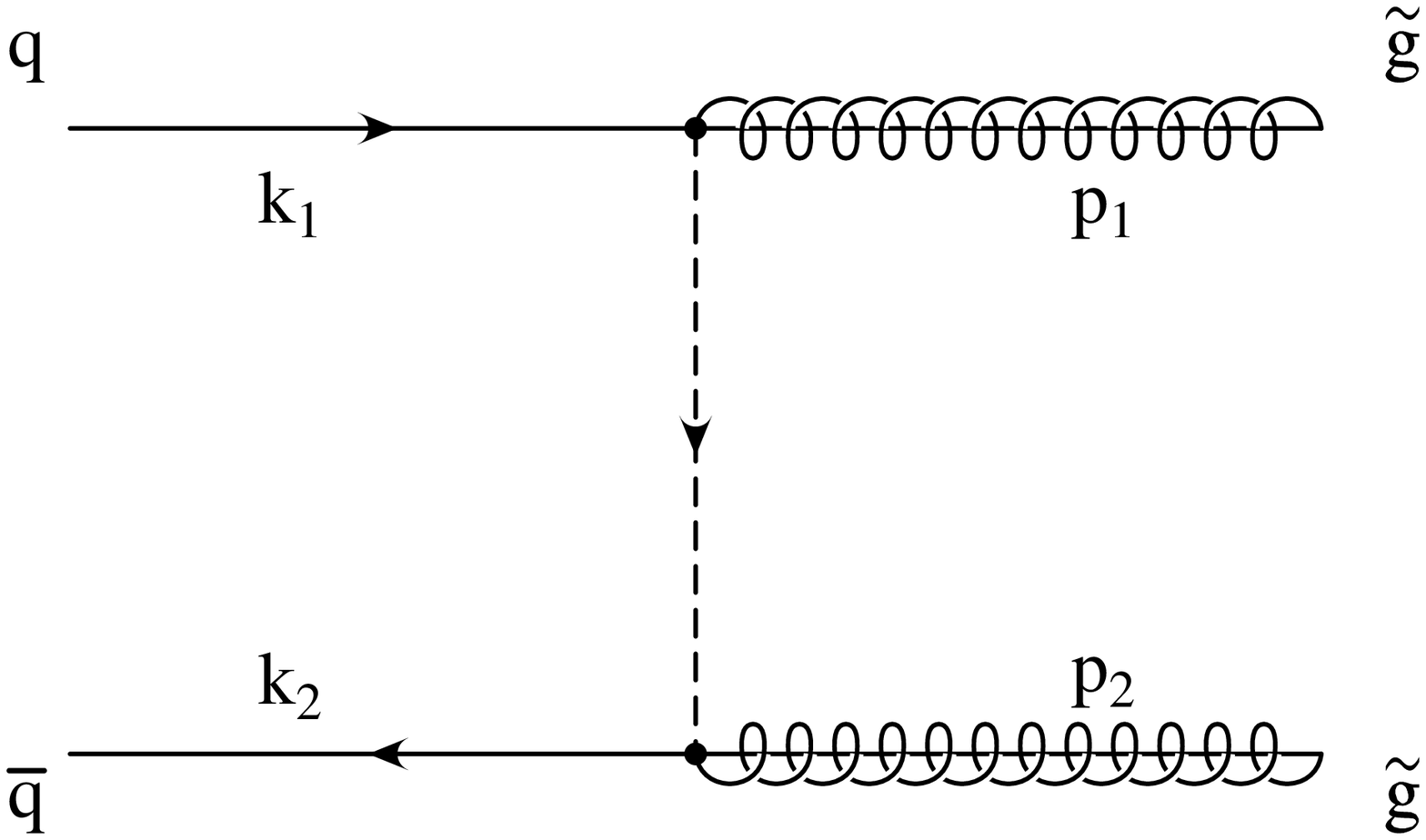}
\includegraphics[scale=0.15]{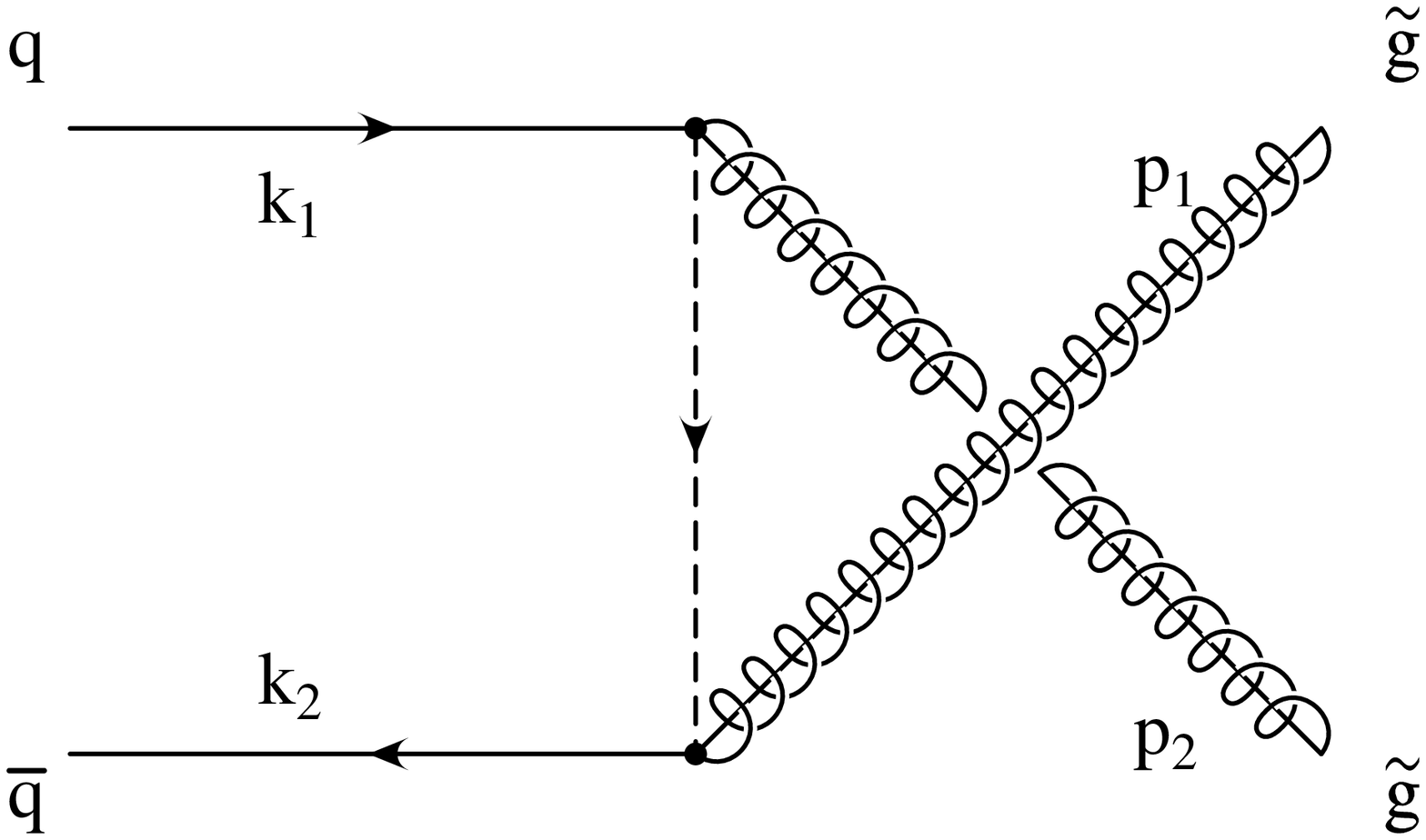}
\\(a)\\
\includegraphics[scale=0.14]{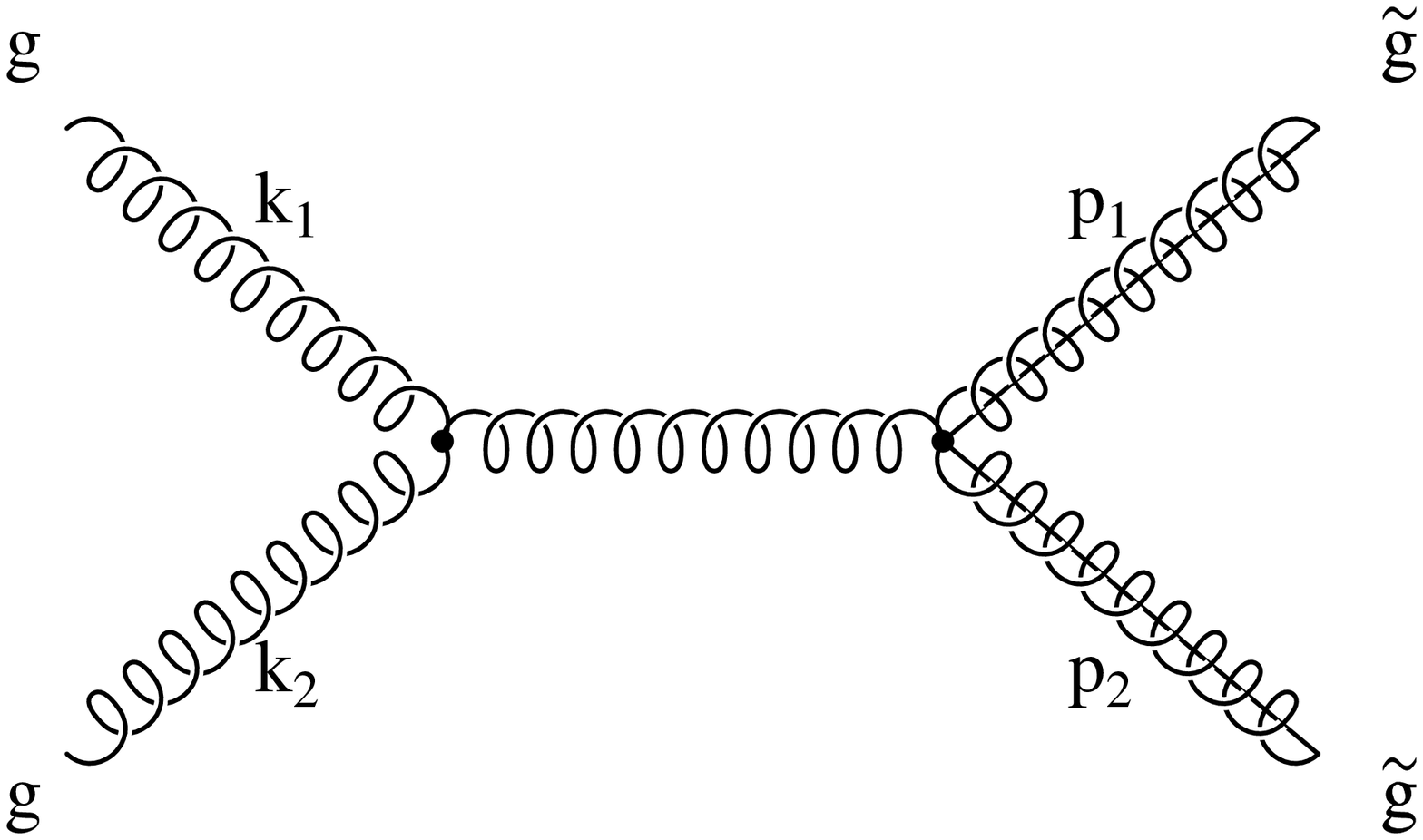}
\includegraphics[scale=0.14]{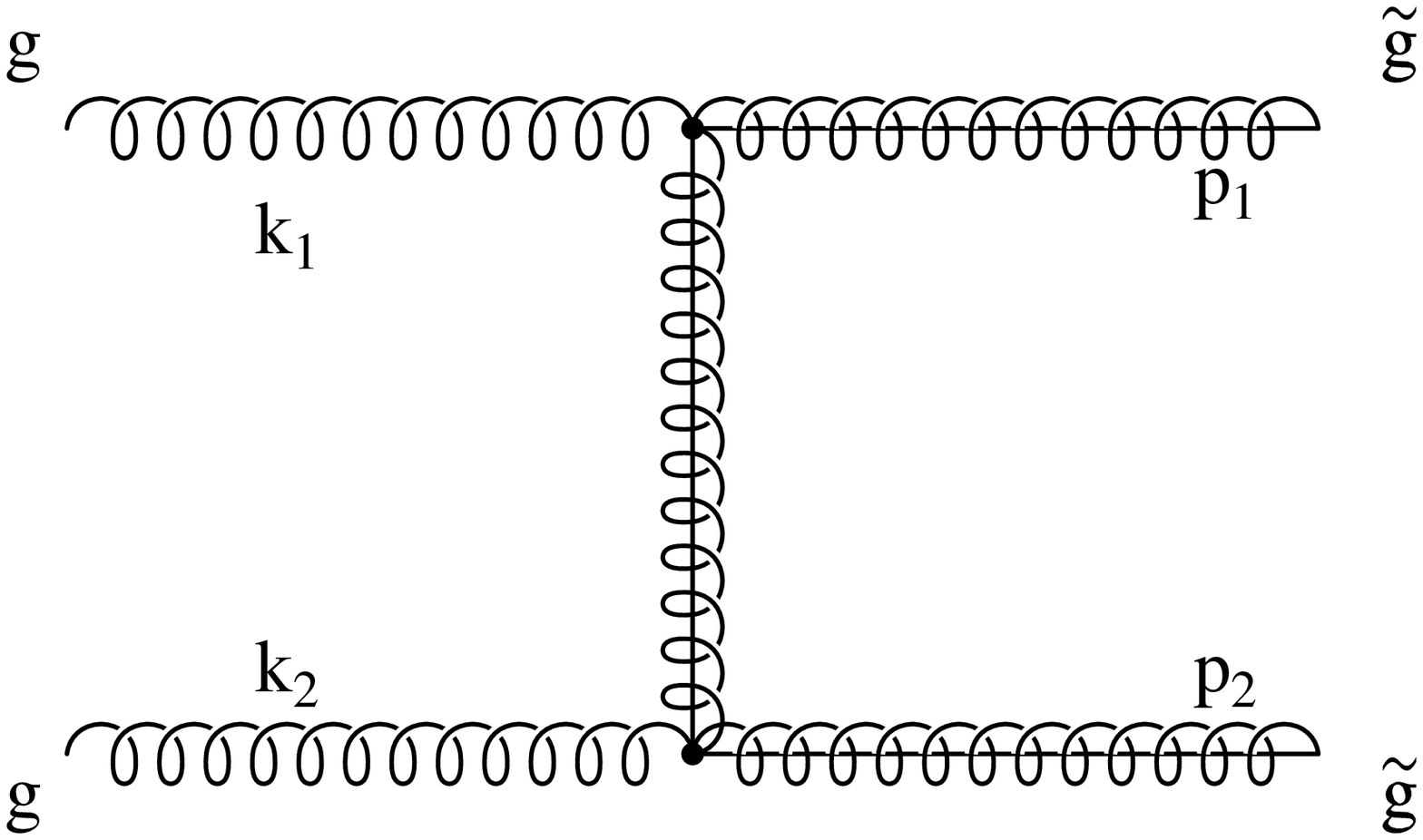}
\includegraphics[scale=0.15]{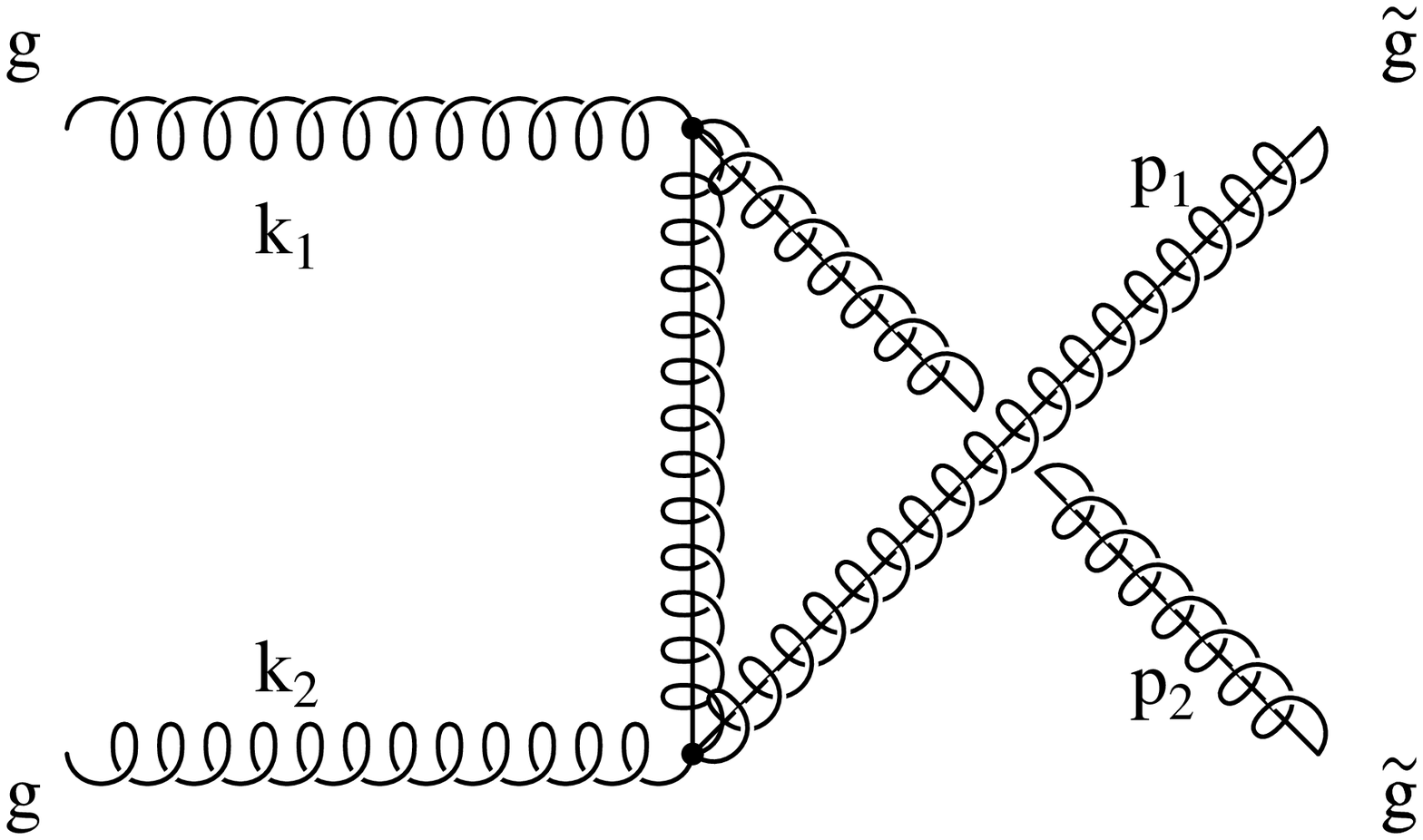}
\\(b)\\
\includegraphics[scale=0.14]{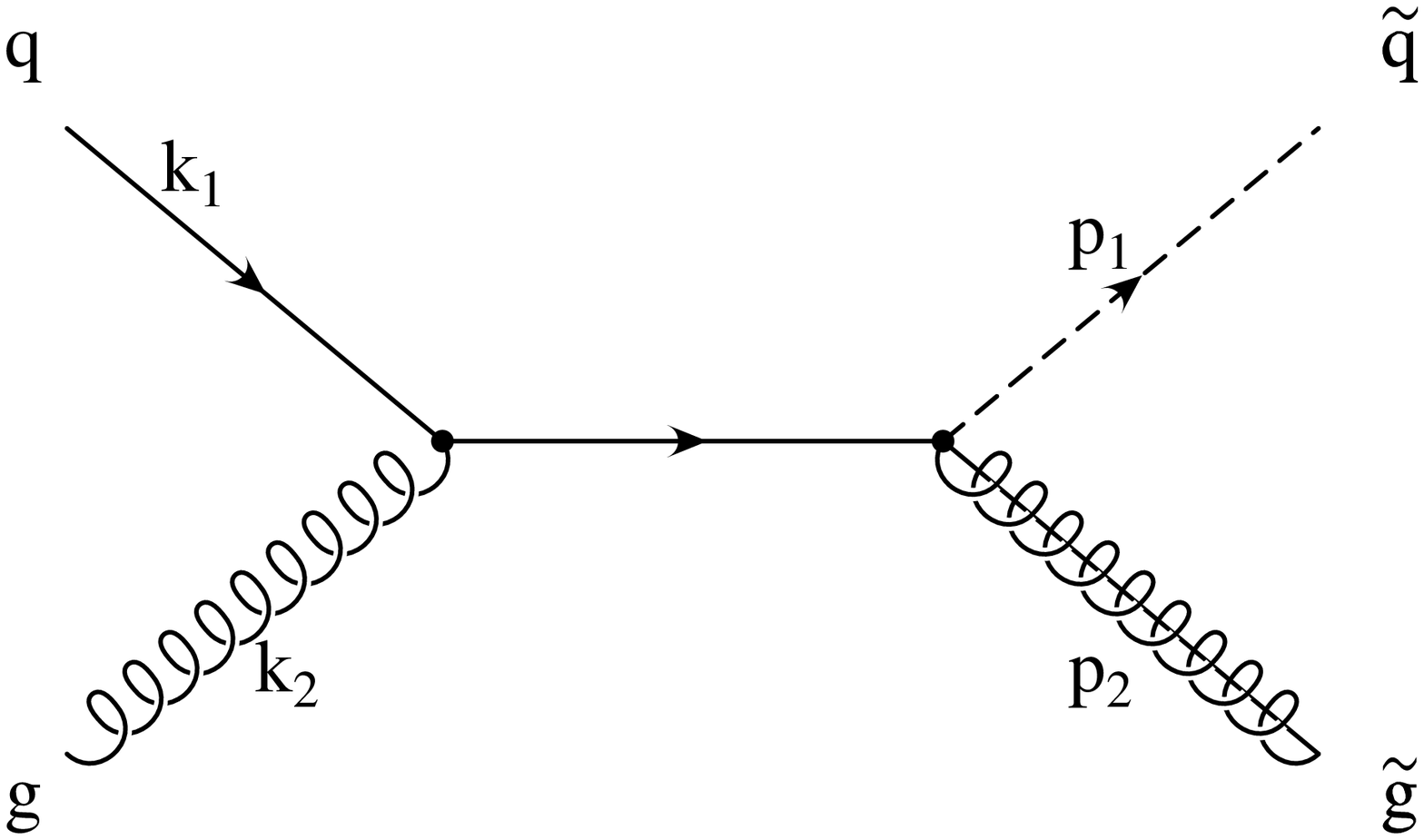}
\includegraphics[scale=0.14]{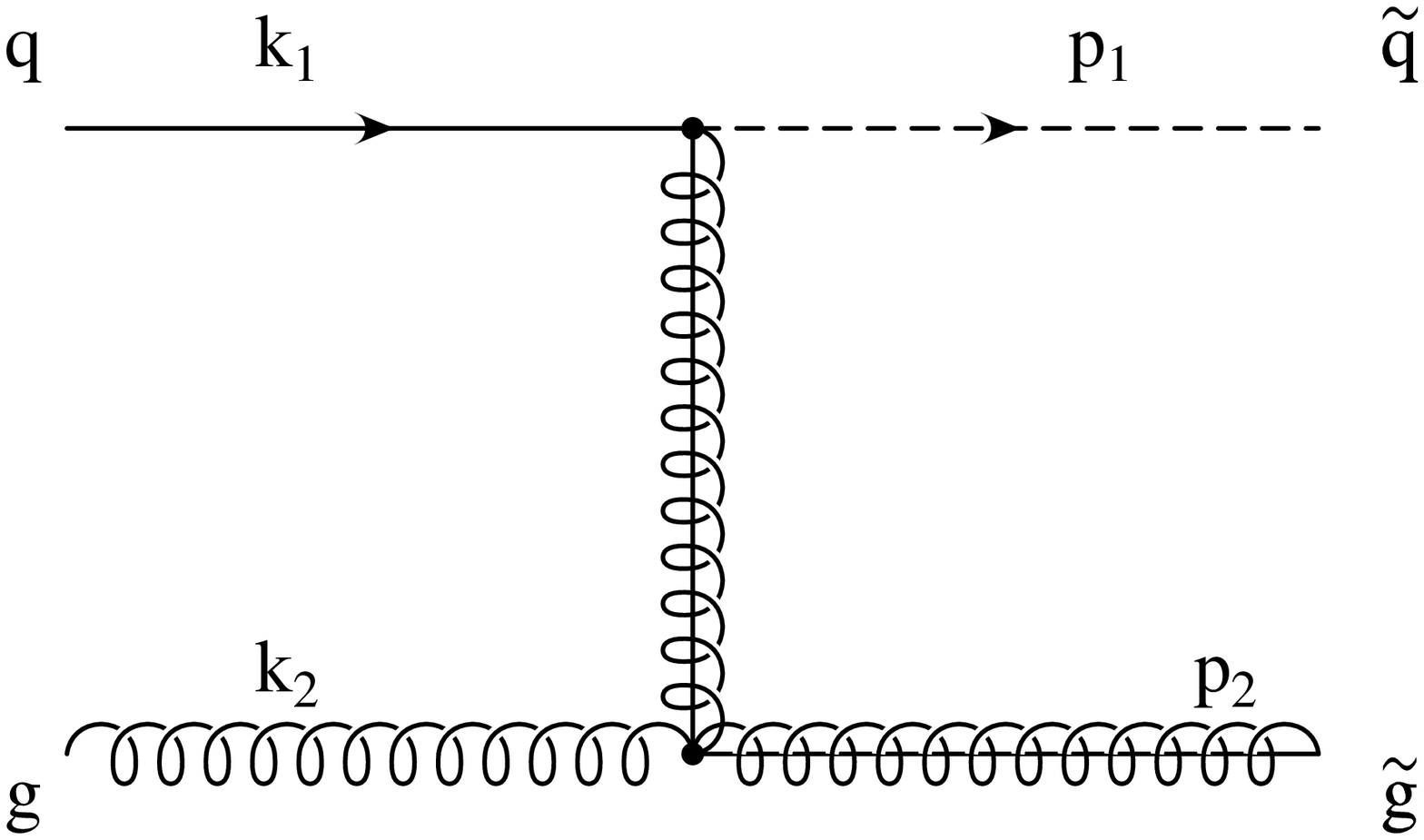}
\includegraphics[scale=0.14]{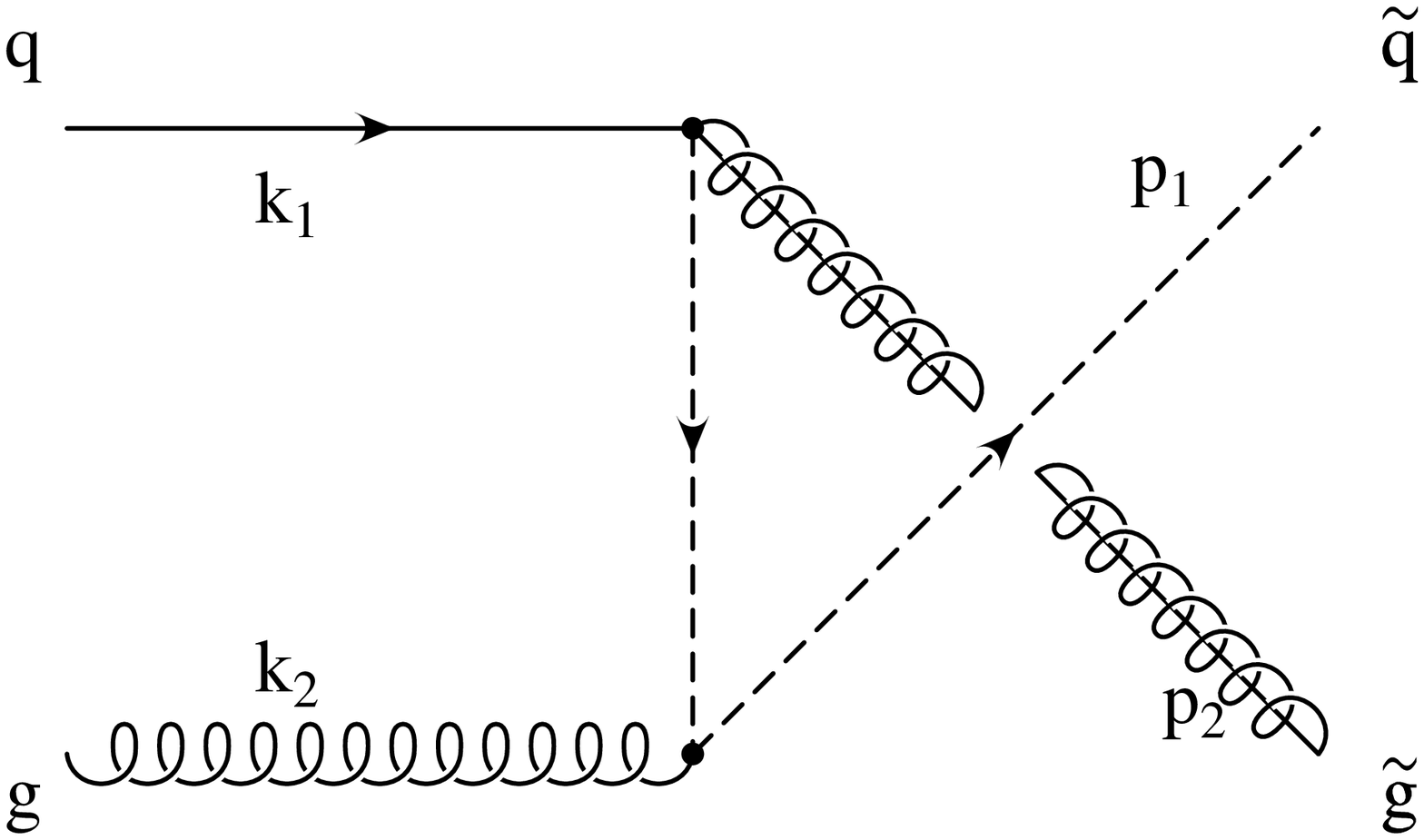}
\\(c)\\
\end{center}
\caption{Feynman diagrams for single (a,b,c) and double (a,b) gluino pair
production.}
 \label{diagrams}
\end{figure}

The LO QCD subprocesses for single gluino production are gluon-gluon and quark-antiquark
anihilation ($g g  \rightarrow \tilde{g} \tilde{g}$ and $q \bar q  \rightarrow \tilde{g} \tilde{g}$),  
and the Compton process $qg\rightarrow \tilde{g}\tilde{q}$, as shown in Fig. \ref{diagrams}. 
For double gluino production only the anihilation processes contribute, obviously. 
These two kinds of events could be separated, in principle, by analysing the
different decay channels for gluinos and squarks \cite{tata,dress}.

Incoming quarks (including incoming $b$ quarks) are assumed to be massless,
such that we have $n_f=5$ light flavours. We only consider final state 
squarks corresponding to the light quark flavours. All 
squark masses are taken equal to $m_{\tilde q}$. We do not consider in detail top squark
production where these assumptions do not hold and which require
a more dedicated treatment~\cite{plehn}.

The invariant cross section for single gluino production can be written as \cite{Dawson}
 \begin{eqnarray}
E\frac{d\sigma }{d^3p}&=& \sum_{ijd} \int_{x_{min}}^1 dx_a
f_i^{(a)}(x_a,\mu )f_j^{(b)}(x_b,\mu ) \nonumber \\
&&\,\,\,\,\,\,\,\,\frac{x_ax_b}{x_a-x_{\perp} \left( \frac{\zeta + \cos \theta}{2\sin \theta} \right) }
\frac{d \hat{\sigma}}{d \hat{t}}(ij\rightarrow \tilde{g} d), 
\end{eqnarray}
where $f_{i,j}$ are the parton distributions of the incoming protons 
and $\frac{d \hat{\sigma}}{d \hat{t}}$ is the LO partonic cross section
\cite{Dawson} for the subprocesses involved.
The identified gluino is produced at center-of-mass angle $\theta$ and 
transverse momentum $p_T$, and $x_{\perp}=\frac{2p_T}{\sqrt{s}}$. The kinematic invariants 
of the partonic reactions $ij\rightarrow \tilde{g}\tilde{g}, \tilde{g}\tilde{q}$ 
are then
\begin{eqnarray}
\hat{s}&=& x_ax_bs, \nonumber \\ 
\hat{t}&=& m_{\tilde{g}}^2-x_ax_{\perp}s\left( \frac{\zeta-\cos \theta}{2\sin \theta}\right), 
\nonumber \\ 
\hat{u}&=& m_{\tilde{g}}^2-x_bx_{\perp}s\left(\frac{\zeta+\cos \theta}{2\sin \theta}\right).
\end{eqnarray}
Here
\begin{eqnarray}
x_b&=&\frac{2\upsilon + x_ax_{\perp}s\left( \frac{\zeta-\cos \theta}{\sin \theta}\right)
}{2x_as-x_{\perp}s\left(\frac{\zeta+\cos \theta}{\sin \theta}\right) },
\nonumber \\ 
x_{min}&=&\frac{2\upsilon + x_{\perp}s\left(\frac{\zeta+\cos \theta}{\sin \theta}\right) }
{2s-x_{\perp}s\left( \frac{\zeta-\cos \theta}{\sin \theta}\right) }, \nonumber \\
\zeta &=& {\left( 1+\frac{4m_{\tilde{g}}^2\sin ^2
\theta}{x_\perp^2s} \right)}^{1/2}, \nonumber \\ 
\upsilon &=& m_d^2-m_{\tilde{g}}^{2}, 
\end{eqnarray}
where $m_{\tilde{g}}$ and $m_d$ are the masses of the final-state partons
produced. 
The center-of-mass angle $\theta$ and the differential cross section above
can be easilly written in terms of the pseudorapidity variable $\eta=-\ln \tan
(\theta/2)$,
 which is one of the experimental observables.
The total cross section for the gluino production can be obtained from above
upon integration.

\begin{figure}[t]
\includegraphics[scale=0.35]{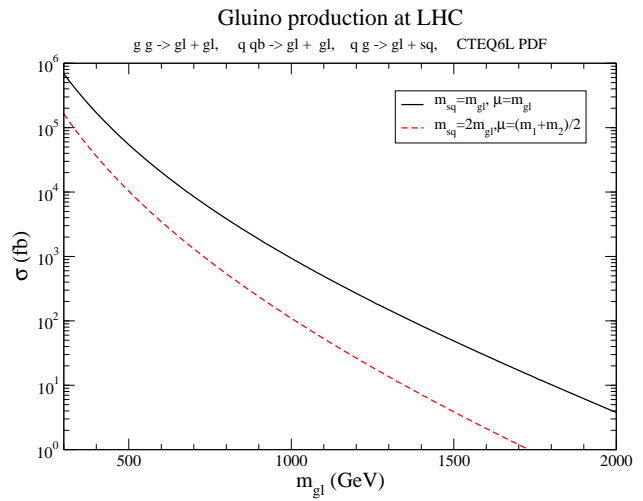}
\caption{The total LO cross section for gluino production at the LHC as a function of the gluino masses. Parton densities: CTEQ6L, with two assumptions on the squark masses and choices of the hard scale.} 
\label{fig:sigtot}
\end{figure}

In Fig.\ref{fig:sigtot} we present the LO QCD total cross section for gluino 
production at the LHC as a function of the gluino masses. 
We use the CTEQ6L \cite{Pumplin:2002vw}, parton densities, with two assumptions on the squark masses and 
choices of the hard scale. 
The results show a strong dependence on the masses of gluinos and
squarks, and also a larger cross section in the degenerated mass case, which
agrees with the results presented at \cite{tata}. 

The search for gluinos and squarks (as well as other searches for SUSY particles) and the
possibility of detecting them will depend on their real masses. We use the SPS 
values from Table I and proceed to the calculation of differential 
distributions for producing gluinos in all presented scenarios. 
From now on we restrict ourselves to the
production of two gluinos, picking only the anihilation processes as explained
above. The calculation
of producing a single gluino (including the Compton process) is done in a more 
detailed publication\cite{BR}. The results obtained will show the possibility 
of discriminating among the different SPS scenarios.

\begin{figure}[t]
\includegraphics[scale=0.35]{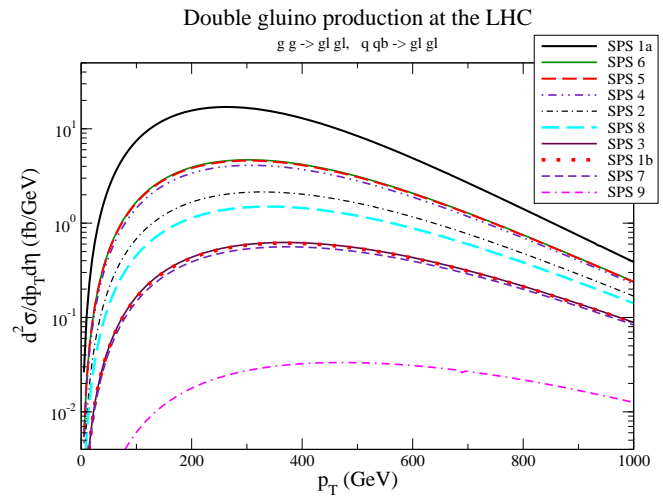}
\caption{The LO $p_T$ distributions for gluino production at the LHC for the different SPS points 
\cite{sps1,sps2}. We use CTEQ6L parton densities, and $\mu^2=m_{\tilde{g}}^2+p_T^2$ as a hard scale.} 
\label{fig:sigpt}
\end{figure}

\begin{figure}[t]
\includegraphics[scale=0.35]{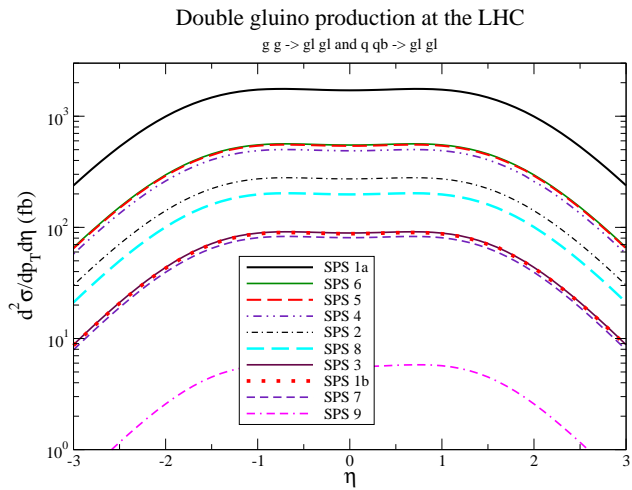}
\caption{The LO pseudorapidity distributions for gluino production at the LHC for the different 
SPS points \cite{sps1,sps2}. We use CTEQ6L parton densities, and $\mu^2=m_{\tilde{g}}^2+p_T^2$ as a hard scale.}
\label{fig:sigeta}
\end{figure}

In Figs.\ref{fig:sigpt} and \ref{fig:sigeta} we present the transverse momentum and pseudorapidity distributions for double
gluino production at LHC energies. 
The results show a similar behavior of the $p_T$ and $\eta$ 
dependencies in all scenarios, 
but a huge diference in the magnitude for different scenarios - 
 SPS1a gives the bigger values, SPS9 the smallest one. Also, we find very close values for 
 SPS1b, SPS3 (mSUGRA) and SPS7 (GMSB), which makes difficult to discriminate
 between these mSUGRA and GMSB models. The same occurs for SPS5 and SPS6 
 (both mSUGRA).

To conclude, we have investigated gluino production at the LHC, which might discover supersymmetry 
over the next years. Gluinos are color octet fermions and play a major role to understanding sQCD. 
Because of their large mass as predicted in several scenarios, up to now the LHC is the only possible 
machine where they could be found.
 
Regarding the strong sector, the Feynman rules are the same for both MSSM and SUSYLR models. 
Therefore, our results for gluino production are equal in both models.
Besides, our results depend on the gluino and squark masses and no other SUSY parameters.
Since the masses of gluinos come only from the soft terms, measuring their masses can test the soft
SUSY breaking approximations. We have considered all the SPS scenarios and showed the corresponding
differences on the magnitude of the production cross sections. From this it is easy to distinguish 
mAMSB from the other scenarios. However, it is not so easy to distinguish mSUGRA from GMSB 
depending on the real values of masses of gluinos and squarks (if SPS1b and SPS7, provided the gluino
and squark masses are almost similar in these two cases). For the other cases, such discrimination
can be done.

\begin{acknowledgments}
This work was partially financed by the Brazilian funding agency
CNPq, CBM under contract number 472850/2006-7, and MCR under contract number 
309564/2006-9.
\end{acknowledgments}

\end{document}